\documentclass[aps,onecolumn,showpacs,footinbib]{revtex4}%
\usepackage[latin9]{inputenc}
\usepackage{amssymb}
\usepackage{amsmath}
\usepackage{amscd}
\usepackage{graphicx}
\usepackage{subfigure}
\usepackage{float}
\begin{document}
\draft
\title{Jaynes-Cummings model with a collective atomic mode}
\author{Shi-Biao Zheng\thanks{%
E-mail: sbzheng@pub5.fz.fj.cn}}
\address{Department of Electronic Science and Applied Physics\\
Fuzhou University\\
Fuzhou 350002, P. R. China}
\date{\today }

\begin{abstract}
We study the dynamics of a single control atom and an atomic sample
interacting with a nonresonant cavity mode. The control atom is driven by an
auxiliary classical field. Under certain conditions, the coherent energy
exchange between the control atom and the atomic sample induced by the
cavity mode is described by the Jaynes-Cummings model. The idea provides a
possibility for quantum state engineering and reconstruction for collective
atomic modes.
\end{abstract}


\vskip 0.5cm \maketitle

\narrowtext

The Jaynes-Cummings model (JCM) [1,2], which describes the interaction of a
two-level atom and a quantized electromagnetic field, is a cornerstone for
treatment of the interaction between light and matter in quantum optics. It
gives arise to many quantum phenomena that can not be explained in classical
terms, such as the collapses and revivals of the atomic population inversion
[3], squeezing of the field [4], and atom-cavity entanglement [5]. Recent
experiments with Rydberg atoms and microwave photons in a superconducting
cavity have turned the JCM from a theoretical curiosity to a useful and
testable enterprise [6]. Such a system is also suitable for quantum state
engineering and quantum information processing. Up to now, Fock states [7],
Schrodinger cat states [8-10], and entangled states [11,12] have been
produced in cavity quantum electrodynamics (QED) experiments. The quantum
logic gate using a Rydberg atom and a cavity field as quantum bits has also
been experimentally realized [13].

In the past few years, there has been interest in the implementation of the
Jaynes-Cummings Hamiltonian with the quantized field replaced by other
bosonic systems. A typical example is the laser-assisted coupling between
the internal and external degrees of freedom of a trapped ion, in which the
excitation (deexcitation) of the electronic state is accompanied by the
annihilation (creaction) of a phonon in the vibrational mode [14]. Thanks to
the advances in ion trapping and laser cooling, various quantum states and
quantum information processsors have been experimentally implemented with
trapped ions [15-18]. Recently, Molmer has shown that the dynamics of a pair
of Bose-Einstein condensates can be described by the Jaynes-Cummings
Hamiltonian [19].

In this paper, we study the dynamics of a system consisting of a single
control atom and an N-atom sample, both of which are couple to a common
cavity mode. Meanwhile, the control atom is driven by an auxiliary clasiscal
field. Through suitable dispersive interactions, we show that the dynamics
of the coupled control atom and atomic ensemble can be described by an
effective Jaynes-Cummings model, in which the collective ensemble atomic
spin is treated as a bosonic mode. One can switch from resonant JCM to
nonresonant JCM by tuning the Rabi frequency of the classical field. Through
suitable choices of interaction strengths and interaction times, one may in
principle generate various quantum states of the collective atomic mode. The
idea can be generalized to generate entangled states for two or more atomic
samples.

We assume that each atom has one excited state $\left| e\right\rangle $ and
one ground state $\left| g\right\rangle $. The Hamiltonian for the whole
system is (assuming $\hbar =1$)
\begin{equation}
H=H_0+H_i,
\end{equation}
where

\begin{equation}
H_0=\omega _ca^{+}a+\omega _0(S_{z,c}+\sum_{j=1}^NS_{z,j}),
\end{equation}
\begin{equation}
H_i=\Omega (S_c^{-}e^{i\omega _dt}+S_c^{+}e^{-i\omega
_dt})+g(a^{+}S_c^{-}+aS_c^{+})+g\sum_{j=1}^N(a^{+}S_j^{-}+aS_j^{+}).
\end{equation}
$S_{z,c}=\frac 12(\left| e_c\right\rangle \left\langle e_c\right| -\left|
g_c\right\rangle \left\langle g_c\right| )$, $S_c^{+}=\left|
e_c\right\rangle \left\langle g_c\right| $, and $S_c^{-}=\left|
g_c\right\rangle \left\langle e_c\right| $ are the inversion, rising, and
lowering operators for the control atom, $S_{z,j}$, $S_j^{-}$, and $S_j^{+}$
denote the inversion, rising, and lowering operators for the jth atom in the
atomic sample, $a^{+}$ and $a$ are the creation and annihilation operators
for the cavity mode, $\omega _0$ is the atomic transition frequency, $\omega
_c$ is the cavity frequency, $\omega _d$ is the frequency of the classical
field, g is the atom-cavity coupling strength, and $\Omega $ is the Rabi
frequency of the classical field. Under the condition $\delta _c=\omega
_0-\omega _c\gg g\sqrt{N(\stackrel{-}{n}+1)}$, with $\stackrel{-}{n}$ being
the mean photon number of the cavity field, there is no energy exchange
between the atomic system and the cavity. The dispersive atom-cavity
interaction leads to photon-number dependent Stark shifts and dipole
couplings for the atomic system. In the case $\delta _d=\omega _0-\omega
_d\gg \Omega $, the classical field only induces a Stark shift. Then the
effective Hamiltonian is [20]

\begin{eqnarray}
H_e &=&\lambda _d(\left| e_c\right\rangle \left\langle e_c\right| -\left|
g_c\right\rangle \left\langle g_c\right| )+\lambda _c(\left|
e_c\right\rangle \left\langle e_c\right| aa^{+}-\left| g_c\right\rangle
\left\langle g_c\right| a^{+}a) \\
&&+\lambda _c\sum_{j=1}^N(\left| e_j\right\rangle \left\langle e_j\right|
aa^{+}-\left| g_j\right\rangle \left\langle g_j\right| a^{+}a)+\lambda
_c\sum_j^N(S_c^{+}S_j^{-}+S_c^{-}S_j^{+})+\sum_{j,k=1}^NS_j^{+}S_k^{-}],%
\text{ }  \nonumber \\
j &\neq &k,  \nonumber
\end{eqnarray}
where $\lambda _d=\Omega ^2/\delta _d$ and $\lambda _c=g^2/\delta _c$.

Since $[a^{+}a,H_e]=0$, the photon-number conserves during the interaction.
If the cavity is initially in the vacuum state it will remain in this state
and the effective Hamiltonian reduces to
\begin{eqnarray}
H_e &=&\lambda _d(\left| e_c\right\rangle \left\langle e_c\right| -\left|
g_c\right\rangle \left\langle g_c\right| )+\lambda _c\left| e_c\right\rangle
\left\langle e_c\right| \\
&&+\lambda _c\sum_j^N(S^{+}S_j^{-}+S^{-}S_j^{+})+\lambda
_c\sum_{j,k=1}^NS_j^{+}S_k^{-},  \nonumber \\
\text{ }j &\neq &k.
\end{eqnarray}
Setting
\begin{eqnarray}
b &=&\frac 1{\sqrt{N}}\sum_{j=1}^NS_j^{-}, \\
b^{+} &=&\frac 1{\sqrt{N}}\sum_{j=1}^NS_j^{+},  \nonumber \\
n_b &=&\sum_{j=1}^N\left| e_j\right\rangle \left\langle e_j\right| ,
\nonumber
\end{eqnarray}
we have
\begin{eqnarray}
\left[ b,b^{+}\right] &=&1-\frac 2Nn_b, \\
\left[ n_b,b^{+}\right] &=&b^{+},  \nonumber \\
\left[ n_b,b\right] &=&-b.  \nonumber
\end{eqnarray}
Suppose that $N\gg 1,\stackrel{-}{n}_b$, with $\stackrel{-}{n}_b$ being the
average excitation number of the atomic sample. Then $b$, $b^{+}$ can be
regarded as the bosonic operators and the atomic sample can be regarded as a
bosonic system. The Hamiltonian can be rewritten as
\begin{eqnarray}
H_e &=&(2\lambda _d+\lambda _c)S_{z,c}+\sqrt{N}\varepsilon b^{+}b \\
&&\ +\varepsilon (S_c^{+}b+S_c^{-}b^{+}),\text{ }  \nonumber
\end{eqnarray}
where $\varepsilon =\sqrt{N}\lambda _c$. Here we have discarded the constant
energy $\lambda _c/2$. The Hamiltonian $H_e$, showing complete analogy with
the Jaynes-Cummings Hamiltonian, describes the oscillatory exchange of an
excitation between the control atom and collective atomic mode.

Under the condition $2\lambda _d=(N-1)\lambda _c$, the Hamiltonian $H_e$,
describing the resonant coupling between the control atom and the atomic
mode, leads to the transition

\begin{eqnarray}
\left| e_c\right\rangle \left| n\right\rangle &\rightarrow &e^{-i\sqrt{N}%
\varepsilon t}[\cos (\varepsilon t)\left| e_c\right\rangle \left|
n\right\rangle -i\sin (\varepsilon t)\left| g_c\right\rangle \left|
n+1\right\rangle ], \\
\left| g\right\rangle \left| n+1\right\rangle &\rightarrow &e^{-i\sqrt{N}%
\varepsilon t}[\cos (\varepsilon t)\left| g_c\right\rangle \left|
n+1\right\rangle -i\sin (\varepsilon t)\left| e_c\right\rangle \left|
n\right\rangle ],  \nonumber
\end{eqnarray}
where $\left| n\right\rangle $ denotes the Fock state for the atomic mode.
The dynamics provides a possibility for realizing various intersting
phenomena for the atomic matter wave field. The coupling also makes it
possible to engineer particular quantum states of the atomic sample. Suppose
that the system is intially in the state $\left| e_c\right\rangle \left|
0\right\rangle $. After an interaction time $t_1=\pi /2\varepsilon $, the
collective atomic mode evolves to the Fock state $\left| 1\right\rangle $,
with the control atom left in the ground state $\left| g_c\right\rangle $.
We then drive the control atom to the excited state $\left| e_c\right\rangle
$ via a classical field. After a second interaction time $t_2=\pi /2\sqrt{2}%
\varepsilon $, the collective atomic mode evolves to the two-excitation Fock
state $\left| 2\right\rangle $. Repeating the procedure we can generate Fock
states for the collective atomic mode with larger excitation-numbers.

The coupling can also be used to generate Schrodinger cat states for the
collective atomic mode. Suppose that the control atom is initially in the
ground state $\left| g_c\right\rangle $ and the atomic sample initially in
the coherent state $\left| \alpha \right\rangle $ ($\alpha =\left| \alpha
\right| e^{i\theta }$). Under the conditions $\left| \alpha \right| \gg 1$
and $gt\ll 4\stackrel{-}{n}$ ($\stackrel{-}{n}=\left| \alpha \right| ^2$)
the interaction leads to [21]
\begin{equation}
\left| \psi (t)\right\rangle =\frac 1{\sqrt{2}}[e^{-i\sqrt{\stackrel{-}{n}}%
\varepsilon t/2}\left| \alpha ^{+}(t)\right\rangle \left| \phi
_a^{+}(t)\right\rangle -e^{-i\sqrt{\stackrel{-}{n}}\varepsilon t/2}\left|
\alpha ^{+}(t)\right\rangle \left| \phi _a^{+}(t)\right\rangle ],
\end{equation}
where
\begin{equation}
\left| \alpha ^{\pm }(t)\right\rangle =e^{-\stackrel{-}{n}/2}\sum_ne^{\pm (n-%
\stackrel{-}{n})^2\varepsilon t/8\stackrel{-}{n}^{3/2}}\frac{(\alpha e^{-i%
\sqrt{N}\varepsilon t}e^{\mp i\varepsilon t/2\sqrt{\stackrel{-}{n}}})^n}{%
\sqrt{n!}},
\end{equation}
\begin{equation}
\left| \phi _c^{\pm }(t)\right\rangle =\frac 1{\sqrt{2}}(e^{i\theta }e^{\mp
i\varepsilon t/2\sqrt{\stackrel{-}{n}}}\left| e_c\right\rangle \pm \left|
g_c\right\rangle ).
\end{equation}
Thus the collective atomic mode splits into two quasicoherent components
with different phases correlated with the states of the control atom. As
shown in the Jaynes-Cummings dynamics with a cavity field [22], the
mesoscopic coherence and decoherence of the collective atomic mode can be
revealed by the collapses and revivals of the Rabi oscillations. We can
reconstruct the quantum states of the collective atomic mode by detecting
the state of the control atom after the resonant Jaynes-Cummings dynamics
[23,24].

Set $\delta =2\lambda _d-(N-1)\lambda _c\gg \varepsilon $. Then the
probability that the control atom exchanges energy with the atomic sample is
negligible and the effective Hamiltonian reduces to [8]
\begin{equation}
H_e=\frac{\varepsilon ^2}\delta (\left| e_c\right\rangle \left\langle
e_c\right| bb^{+}-\left| e_c\right\rangle \left\langle e_c\right| b^{+}b).
\end{equation}
Suppose that the atom is initially in the superposition state $\frac 1{\sqrt{%
2}}(\left| e_c\right\rangle +\left| g_c\right\rangle )$ and the atomic
sample initially in the coherent state $\left| \alpha \right\rangle $. The
effective Hamiltonian gives arise to the Schr\"odinger cat state
\begin{equation}
\left| \psi (t)\right\rangle =\frac 1{\sqrt{2}}[e^{-i\varepsilon ^2t/\delta
}\left| \alpha e^{-i\varepsilon ^2t/\delta }\right\rangle \left|
e_c\right\rangle +\left| \alpha e^{i\varepsilon ^2t/\delta }\right\rangle
\left| g_c\right\rangle ].
\end{equation}
The effective Hamiltonian can also be used to directly measure the Wigner
function of the atomic matter wave field [25].

The idea can be used to generate entangled states for two atomic samples.
Suppose that a control atom and two atomic samples dispersively interact
with the vacuum cavity. Meanwhile, the control atom is driven by a
nonresonant classical field. Under the above mentioned conditions, the
effective Hamiltonian is
\begin{eqnarray}
H_e &=&(2\lambda _d+\lambda _c)S_{z,c}+\sqrt{N}\varepsilon
(b_1^{+}b_1+b_2^{+}b_2+b_1^{+}b_2+b_2^{+}b_1) \\
&&\ \ +\varepsilon [S_c^{+}(b_1+b_2)+S_c^{-}(b_1^{+}+b_2^{+})],\text{ }
\nonumber
\end{eqnarray}
where $b_1$ and $b_2$ represent the annihilation operators for the two
collective atomic modes, respectively. Suppose that the control atom is
initially in the excited state $\left| e_c\right\rangle $ and the two atomic
samples initially in the vacuum state $\left| 0_1\right\rangle \left|
0_2\right\rangle $. In the case that $2\lambda _d=(2N-1)\lambda _c$, the
evolution of the system is

\begin{equation}
\left| \psi (t)\right\rangle =e^{-i2\sqrt{N}\varepsilon t}[\cos (\sqrt{2}%
\varepsilon t)\left| e_c\right\rangle \left| 0_1\right\rangle \left|
0_2\right\rangle -i\sin (\sqrt{2}\varepsilon t)\left| g_c\right\rangle
(\left| 1_0\right\rangle \left| 0_2\right\rangle +\left| 0_0\right\rangle
\left| 1_2\right\rangle )/\sqrt{2}].
\end{equation}
After an interaction time $t=\pi /2\sqrt{2}\varepsilon $, the two collective
atomic modes evolves to the maximally entangled state
\begin{equation}
\left| \psi _m\right\rangle =e^{-i\sqrt{N}\pi /\sqrt{2}}(\left|
1_0\right\rangle \left| 0_2\right\rangle +\left| 0_0\right\rangle \left|
1_2\right\rangle )/\sqrt{2},  \nonumber
\end{equation}
with the control atom left in the ground state $\left| g_c\right\rangle $.

We note that the entanglement operation can be applied to the system
involving n atomic samples. In this case, the effective Hamiltonian is
\begin{eqnarray}
H_e &=&(2\lambda _d+\lambda _c)S_{z,c}+\sqrt{N}\varepsilon
\sum_{j=1}^n\sum_{k=1}^nb_j^{+}b_k \\
&&\ \ \ +\varepsilon (S_c^{+}\sum_{j=1}^nb_j+S_c^{-}\sum_{k=1}^nb_j^{+}).%
\text{ }  \nonumber
\end{eqnarray}
We again assume that the control atom is initially in the excited state $%
\left| e_c\right\rangle $ and the atomic samples initially in the state $%
\left| 0_1\right\rangle \left| 0_2\right\rangle ...\left| 0_n\right\rangle $%
. With the choice $2\lambda _d=(nN-1)\lambda _c$, we obtain the state
evolution

\begin{eqnarray*}
\left| \psi (t)\right\rangle &=&e^{-in\sqrt{N}\varepsilon t}[\cos (\sqrt{n}%
\varepsilon t)\left| e_c\right\rangle \left| 0_1\right\rangle \left|
0_2\right\rangle ...\left| 0_n\right\rangle \\
&&-i\sin (\sqrt{n}\varepsilon t)\left| g_c\right\rangle (\left|
1_0\right\rangle \left| 0_2\right\rangle ...\left| 0_n\right\rangle +\left|
0_0\right\rangle \left| 1_2\right\rangle \left| 0_3\right\rangle ...\left|
0_n\right\rangle +...+\left| 0_0\right\rangle \left| 0_2\right\rangle
...\left| 0_3\right\rangle \left| 1_n\right\rangle )/\sqrt{n}].
\end{eqnarray*}
After an interaction time $t=\pi /2\sqrt{n}\varepsilon $, the atomic modes
evolves to
\begin{equation}
\left| \psi _m(t)\right\rangle =e^{-i\sqrt{nN}\pi /2}(\left|
1_0\right\rangle \left| 0_2\right\rangle ...\left| 0_n\right\rangle +\left|
0_0\right\rangle \left| 1_2\right\rangle \left| 0_3\right\rangle ...\left|
0_n\right\rangle +...+\left| 0_0\right\rangle \left| 0_2\right\rangle
...\left| 0_3\right\rangle \left| 1_n\right\rangle )/\sqrt{n}.  \nonumber
\end{equation}
This is a W state [26], whose entanglement is robust against qubit loss,
global dephasing, and qubit flip noise. Due to the robustness multiqubit W
states might lead to stronger nonclassicality [27] than the
Greeberger-Horne-Zeilinger states [28] and be useful in quantum information
processing [29].

It is necessary to address the experimental implementation of the proposed
model. In recent cavity QED experiments [30,31], Cs atoms were trapped in an
optical cavity, and the $6S_{1/2}$, $F=4\rightarrow 6P_{3/2}$, $F=4$
transition was coupled to the cavity mode. The corresponding coupling
strength is $g=2\pi \times 34MHz$. The decay rates for the atomic excited
state and the cavity mode are $\Gamma =2\pi \times 2.6MHz$ and $\kappa =2\pi
\times 4.1MHz$, respectively. The decoherence rate of the atomic sample due
to atomic spontaneous emission is given by the single-atom spontaneous
emission rate [32]. In order to suppress the influence of the atomic
spontaneous emission one should use Raman coupling. Suppose that two ground
states are coupled to the excited state through the cavity mode of coupling $%
g$ and a classical field of coupling $\alpha $. The cavity mode and
classical field are detuned from the respective transitions by the amounts $%
\Delta +\delta $ and $\Delta $, respectively. Under the condition $\Delta
\gg g,\alpha ,\delta $ the excited state can be adiabatically eliminated and
the two ground states are coupled to the cavity mode through Raman process.
The Raman coupling strength is $g^{^{\prime }}=\frac{g\alpha }2(\frac 1{%
\Delta +\delta }+\frac 1\Delta )$ [33]. Suppose that $\delta \gg g^{^{\prime
}}$. Then the Raman coupling is far off-resonant and the cavity mode is
virtually excited. This leads to the coupling between the control atom and
the atomic sample. The entire process is a four-photon transition [34].
Meanwhile, the energy difference between the two ground states of the
control atom can be modified through an additional classical field. Set $%
N=10^4$, $\Delta =100g$, $\delta =10g$, and $\alpha =g$. Then the coupling
strength between the control atom and the collective atomic mode is $%
\varepsilon =\sqrt{N}g^{^{\prime }2}/\delta \simeq 2\pi \times 3.1\times
10^4Hz$. Both the atomic system and the cavity mode are virtually excited.
The effective decoherence rates due to the atomic spontaneous emission and
cavity decay are $\Gamma ^{^{\prime }}=\Gamma g^2/\Delta ^2=$ $2\pi \times
260Hz$ and $\kappa ^{^{\prime }}=\kappa g^{^{\prime }2}/\delta ^2\simeq $ $%
2\pi \times 3.7Hz$, respectively. In this case both the effective
decoherence rates $\Gamma ^{^{\prime }}$ and $\kappa ^{^{\prime }}$ are much
smaller than the coupling strength $\varepsilon $. For the generation of the
Fock state $\left| 1\right\rangle $, the required interaction time is $%
t_1=\pi /2\varepsilon \simeq 8.1\mu s$. The infidelity induced by the
decoherence is on the order of $(\Gamma ^{^{\prime }}+\kappa ^{^{\prime
}})t_1\simeq 1.3\times 10^{-2}$.

It should be noted that the derived Jaynes-Cummmings dynamics is valid in
the Lamb-Dicke regime. Recently, a single $^{40}Ca^{+}$ ions was localized
at a fixed position in the cavity with a high precision for a long time
[35]. In a more recent experiment [36], the localization to the Lamb-dicke
limit of the axial motion was demonstrated for a single Cs atom trapped in
an optical cavity. In order to couple only the control atom to the
additional classical field without coupling the N-atom sample as well one
should put the atomic sample in a lattice-like structure and address
selectively the control atom by a focused external beam.

In conclusion, we have shown how to realize the JCM with the collective
atomic bosonic mode. The coupling between the control atom and the atomic
sample is induced by the nonresonant cavity mode, which is always in the
vacuum state. Thus the evolution of the system is insensitive to cavity
decay. The dynamics provides a possibility for realizing and reconstructing
various quantum states of matter wave oscillators. The idea can also be used
for preparation of entangled states for two atomic samples trapped in two
separate cavities. Suppose that each cavity involves a control atom and an
atomic sample. In each cavity the control atom and collective atomic mode
are first entangled via the four-photon transition. Then we switch off the
classical fields which induces the four-photon transition. Meanwhile, we
apply an appropriate classical field to the control atom so that it
undergoes a resonant Raman transition and can emit a photon. Various
entangled states for the two collective atomic modes can be produced by
detecting photons leaking out of the cavities [37]. This may allows one to
test Bell's inequalities [38] with two entangled macroscopic objects. Due to
the long coherence lifetime, the matter wave oscillator is suitable for the
storage of quantum information. The entanglement between two distant atomic
samples has application in quantum communication [32]. Two atomic samples
located in separate cavities can also be entangled by using a fiber to
connect the cavities [39]. The two control atoms are first entangled via the
exchange of an excitation through the fiber. Then the coupling between the
control atom and atomic sample in each cavity leads to the entanglement
between the two atomic samples. The idea can be generalized to entangle more
distant atomic samples.

This work was supported by the National Natural Science Foundation of China
under Grant No. 10674025 and the Doctoral Foundation of the Ministry of
Education of China under Grant No. 20070386002.

\end{document}